\documentclass[12pt,oneside,reqno]{amsart}

\theoremstyle{definition}

\theoremstyle{remark}

\numberwithin{equation}{section}

\usepackage{amssymb}

\begin{document}


\title{Molecular Brownian motion and \\
invariance group of the Bogolyubov equation}

\author{Yuriy\, E.\, Kuzovlev}
\address{Donetsk Institute for Physics and Technology of NASU,
ul.\,R.\,Luxemburg 72, Donetsk 83114, Ukraine}
\email{kuzovlev@kinetic.ac.donetsk.ua}





\begin{abstract}
Statistics of molecular random walks in a fluid is considered with
the help of Bogolyubov equation for generating functional of
distribution functions. An invariance group of this equation is
found. It results in many exact relations between path probability
distribution of a test particle and its correlations with the fluid.
As the consequence, significant restrictions on possible shape of the
path distribution do arise. In particular, the hypothetical Gaussian
form of long-range asymptotic proves to be forbidden, even (and first
of all) under the Boltzmann-Grad limit. An allowed diffusive
asymptotic possesses power-law long tail (cut off by free flight
length).
\end{abstract}


\maketitle

\baselineskip 24 pt

\markboth{}{}

\section{Introduction}

Random wandering of particles of the matter is mechanism of diffusion
and many other transport processes as well as various noises and
fluctuations. What can be its statistics? Strangely, this important
question never was addressed to rigorous statistical mechanics. At
present, seemingly, the answer is obvious without it: since even in
the Lorentz gas asymptotic of random walks is Gaussian \cite{sin},
all the more it must be the same in a usual fluid. However, one can
notice that such conclusion is founded on instinctive identifying
dynamical independence of events in ``real life'' (concrete
realization of many-particle system) and statistical independence of
events in ``theory'' (statistical ensemble of systems). But already
Krylov \cite{kr} thoroughly explained that the first does not imply
the second \footnote{\,In the probability theory {\it by\,
definition}\, $A$ and $B$ are independent if\, $\,\mathcal{P}(A\wedge
B)=\mathcal{P}(A)\mathcal{P}(B)$ \cite{kol}. Therefore statement
``\,$A$\, and \,$B$\, are independent since have nothing to do with
each other'' \cite{rcc} is wrong: such $A$ and $B$ may both depend on
common conditions so that\, $\,\mathcal{P}(A\wedge B)\neq
\mathcal{P}(A)\mathcal{P}(B)\,$ !}.

This can be understood by the example of gas of $\,N\,$ hard balls in
a box if considering their motion as motion of one ball in
$\,3N$-dimensional billiard \cite{gal} which resembles the Lorentz
gas. Here as well initially one-dimensional  (straight-line) motion
becomes multidimensional. However hardly it is possible  to speak
about ergodicity before all the dimensions and scatterers become
apparent, that is before time $\,\sim\frac 12 N(N-1)/(N/\tau )\sim
N\tau\,\sim \,\Omega/\pi r_0^2v_0\,$ passed after start of
observations, with\, $\,\Omega\,$,\, $\,\tau\sim (\pi r_0^2\nu v_0
)^{-1}\,$,\, $\,r_0\,$,\, $\,v_0\,$\, and $\,\nu =N/\Omega\,$\, being
volume of the box, typical free path time of gas particles, radius of
their interaction (diameter of the balls), their characteristic
velocity and mean gas density, respectively. Even for $\,1\,$cm$^3\,$
of the air this is time greater than 1000 years! Thus, it was true
remark \cite{rar} that role of ergodicity in physics is strongly
exaggerated since in physical reality limit $\,N\rightarrow \infty\,$
precedes limit $\,t/\tau \rightarrow\infty \,$.

If it is so then trajectory of test particle in gas of sufficiently
many particles (formally, under $\,N\rightarrow \infty\,$) is
non-ergodic:\, its even very relatively distant fragments (separated
by time intervals $\,\gg \tau\,$) are statistically dependent one on
another although independent in dynamical sense. This means that any
concrete realization of random walk of the test particle
(corresponding to some concrete phase trajectory of the whole system)
possesses its own specific {\it kinetic characteristics} (diffusivity
and mobility, etc.) or, better to say, has no certain kinetic
characteristics \cite{bk3,i1}.

Theoretical tools for investigations of such statistics a long time
ago were presented by N.\,Bogolyubov \cite{bog} and his followers.
That are the BBGKY (Bogolyubov-Born-Green-Kirkwood-Yvon) hierarchy of
equations or equivalent Bogolyubov's equation for generating
functional of many-particle distribution functions. But,
unfortunately, nobody has learned honest use of them without some
cutting off the hierarchy and justifying such vivisection by
imaginary intuitive ``independencies''. Therefore complementary tools
are not unnecessary.

In the present work we describe an invariance group of the Bogolyubov
equation and its solutions. For simplicity, it is formulated directly
with reference to the problem about random walk of test particle in
thermodynamically equilibrium fluid. We want to demonstrate that
exact relations of the group point to essentially non-Gaussian
statistical features of long-range asymptotic of the random walk.
Importantly, in case of gas these relations and their consequences,
along with the mentioned characteristic time scale $\,\sim N\tau\,$,
are indifferent to gas density, even in the limit of dilute gas (or
the Boltzmann-Grad limit).

\section{Equations of molecular random walk}

Let box with volume $\,\Omega\,$ contains $\,N\gg 1\,$ identical
atoms plus one more test particle.  Atoms have mass $\,m\,$,
coordinates $\,{\bf r}_j\,$ and momenta $\,{\bf p}_j\,$
($\,j=1,2...\,N\,$) and interact  with each other via potential
$\,\Phi_a({\bf r}_j-{\bf r}_k)\,$. The test particle has mass  $\,M\,$,
coordinate $\,{\bf R}\,$, momentum $\,{\bf P}\,$ and interacts with
atoms via potential $\,\Phi_b({\bf r}_j-{\bf R})\,$. The potentials are
spherically symmetric and short-range with impenetrable point
core.  Because of interactions the test particle is in chaotic motion,
therefore let us name it ``molecular Brownian particle'' (BP).

We are interested in probability distribution of current position
of BP, $\,{\bf R}(t)\,$, under condition that at initial time moment
$\,t=0\,$ it was placed at certainly known position:\,
$\,{\bf R}(0)={\bf R}_0\,$,\, while personal positions of atoms all the
times are unknown. The simplest statistical ensemble what
satisfies this requirement is determined by the Liouville
equation,\, $\,\partial D_N/\partial
t\,=\,[\,H_N\,,D_N\,]\,$,\, for full normalized distribution
function of the system,\, $\,D_N\,$\,, and initial condition
\begin{equation}
D_N(\,t=0\,)\,=\,\frac {\delta ({\bf R}-{\bf
R}_0)\,\,e^{-\,H_N/T}}{\int d{\bf R}\int d{\bf P}\int_1...\int_N
\delta ({\bf R}-{\bf R}_0)\,e^{-\,H_N/T}}\,\,\label{din}
\end{equation}
to it, where $\,H_N\,$ is full Hamiltonian of the system (including
interactions with the box walls) and\, $\,\int_k ...=\int\int
...\,\,d{\bf r}_k\,d{\bf p}_k\,$\,. Evidently, such ensemble differs
from the Gibbs canonic ensemble by initial BP's localization only.
The latter does not prevent us to introduce particular distribution
functions (DF) \, $\,F_n(t)\,=\,\Omega^{n}\int_{n+1} ...\int_N
D_N(t)\,$\, and then go to the thermodynamical limit
($\,N\rightarrow\infty \,$, $\,\Omega \rightarrow\infty \,$,
$\,\nu=N/\Omega=\,$const\,) just as in  \cite{bog}.  The only,
non-principal, difference from \cite{bog} is that all $\,n$-atom DF
under consideration include also BP's variables, i.e. in fact are
$\,(n+1)$-particle DF, therefore their numeration takes beginning at
zero, and in respect to BP's variables all they are normalized in
usual sense of the probability theory. Writing out complete list of
arguments,\, $\,F_n(t)=\,$ $F_n(t,{\bf R}, {\bf r}^{(n)},{\bf P},{\bf
p}^{(n)}|\,{\bf R}_0\,;\nu\,)\,$\,, where\, $\,{\bf r}^{(n)}=\{{\bf
r}_1...\,{\bf r}_n\,\}\,$, $\,{\bf p}^{(n)}=\{{\bf p}_1...\,{\bf
p}_n\,\}\,$. In particular, $\,F_0(t,{\bf R},{\bf P}|\,{\bf
R}_0\,;\nu\,)\,$ describes BP itself, and $\,\int\! d{\bf R}\int\!
d{\bf P}\,\,F_0\,=\,1\,$. What is for the atomic coordinates, in
respect to them all DF are ``normalized to volume'', exactly as in
\cite{bog}. In other words, the ``weakening of correlations'' between
distant particles holds:\, $\,F_n\,\rightarrow\,F_{n-1}\,G_m({\bf
p}_k)\,$, if $\,k$-th particle keeps away from others, where
$\,G_m({\bf p})\,=\,(2\pi Tm)^{-\,3/2}\exp{(-{\bf p}^2/2Tm)}\,$ is
the Maxwell momentum distribution of a particle with mass $\,m\,$.
The full Liouville equation induces the BBGKY equations
\begin{equation}
\frac {\partial F_n}{\partial t}=[\,H_{n}\,,F_n\,]\,+\,\nu\, \frac
{\partial }{\partial {\bf P}}\int_{n+1}\!\! \Phi^{\,\prime}_b({\bf
R}-{\bf r}_{n+1})\,F_{n+1}\,+\,\nu \sum_{j\,=1}^n\,\frac {\partial
}{\partial {\bf p}_j}\int_{n+1}\!\!\Phi^{\,\prime}_a({\bf r}_j-{\bf
r}_{n+1}) \,F_{n+1}\,\label{fn}
\end{equation}
($\,n\,=\,0,\,1,\,\dots\,$)\, with initial conditions
\begin{equation}
\begin{array}{l}
F_n(t=0)\,=\, \delta({\bf R}-{\bf R}_0)\,F_n^{(eq)}({\bf
r}^{(n)}\,|{\bf R};\nu)\,G_M({\bf P})\prod_{j\,=1}^n G_m({\bf
p}_j)\,\,\,,\label{ic}
\end{array}
\end{equation}
where\, $\,H_{n}\,$ is Hamiltonian of subsystem
``$\,n\,$ atoms plus BP\,'',
$\,\Phi^{\,\prime}_{a,\,b}({\bf r})=\nabla\Phi_{a,\,b}({\bf
r})\,$\,,\, and\, $\,F_n^{(eq)}({\bf r}^{(n)}\,|{\bf R};\nu)\,$
are usual thermodynamically equilibrium DF for
$\,n\,$ atom in presence of BP occupying
point $\,{\bf R}\,$.
In principle, that will do for finding
$\,F_0(t,{\bf R},{\bf P}|\,{\bf R}_0\,;\nu\,)\,$
and thus probability distribution of BP's path,\,
$\,\Delta{\bf R}(t)={\bf R}(t)-{\bf R}_0\,$,\,
without any additional assumptions.

Following Bogolyubov, let us combine all our
DF into generating functional (GF)
\begin{equation}
\mathcal{F}\{t,{\bf R},{\bf P},\psi\,|{\bf
R}_0;\nu\}\,=\,F_0\,+\sum_{n\,=1}^{\infty } \frac {\nu^n}{n!}\int_1
...\int_n F_n \,\prod_{j\,=1}^n \psi({\bf r}_j,{\bf
p}_j)\,\,\label{gf}
\end{equation}
and equations (\ref{fn}) into corresponding
``generating equation'' for it:
\begin{eqnarray}
\frac {\partial \mathcal{F}}{\partial t}\,+\,\frac {\bf
P}{M}\cdot\frac {\partial \mathcal{F}}{\partial {\bf
R}}\,=\,\mathcal{\widehat{L}}\left(\psi,\frac {\delta
}{\delta\psi}\right)\,\mathcal{F}\,\,\,,\label{fe}
\end{eqnarray}
where operator $\,\mathcal{\widehat{L}}\,$
is composed by usual and variational derivatives,
\begin{eqnarray}
\mathcal{\widehat{L}}\left(\psi,\frac {\delta
}{\delta\psi}\right)\,=\,-\int_1 \psi(x_1)\,\, \frac {{\bf
p}_1}{m}\cdot\frac {\partial }{\partial {\bf r}_1}\,\frac {\delta
}{\delta \psi(x_1)}\,\,+\,\;\;\;\;\;\;\;\;\;\;\;\;\;
\;\;\;\;\;\;\;\;\;\;\;\;\;\;\; \;\;\;\;\;\;\;\;\;\;
\;\;\;\;\;\;\;\;\;\;\; \label{l}\\
+\,\,\int_1\,\, [\,1+\psi(x_1)\,]\left[\,\Phi_b({\bf R}-{\bf
r}_{1})\,,\frac {\delta }{\delta
\psi(x_1)}\,\right]\,+\,\;\;\;\;\;\;\;\;\;\;\;\;\;\;\;\;\;
\;\;\;\;\;\;\;\;\;\;\;\;\;\nonumber\\
+\,\frac 12 \int_1\int_2\,\, [\,1+\psi(x_1)\,]\,[\,1+\psi(x_2)\,]
\left[\,\Phi_a({\bf r}_1-{\bf r}_2)\,,\frac {\delta^{\,2} }{\delta
\psi(x_1)\,\delta \psi(x_2)}\,\right]\,\,\,,\nonumber
\end{eqnarray}
with\,  $\,x_j\,=\,\{{\bf r}_j,{\bf p}_j\}\,$\,. This is direct
analogue of equation  (7.9) from \cite{bog}. To make it well visible,
notice that $\,\psi(x)=u(x)/\nu\,$, where $\,u(x)\,$ is functional
argument used in  \cite{bog}, and factor
$\,[1+\psi(x_1)]\,[1+\psi(x_2)]\,$ can be replaced by
$\,[\,\psi(x_1)+\psi(x_2)+\psi(x_1)\psi(x_2)\,]\,$ due to identity\,
$\,\int_1\int_2 \,[\,\Phi_a({\bf r}_1-{\bf r}_2)\,, ...\,]=0\,$.\,
Initial condition to equation (\ref{fe}) is obvious:
\begin{eqnarray}
\mathcal{F}\{0,\,{\bf R},{\bf P},\psi\,|\,{\bf R}_0;\nu\}\, =
\,\delta({\bf R}-{\bf R}_0)\,G_M({\bf P})\, \mathcal{F}^{(eq)}\{\phi
|\,{\bf R};\nu\}\,\,\,,\,\,\,\,\,\,\,\,\label{icf}\\
\mathcal{F}^{(eq)}\{\phi |\,{\bf R};\nu\}\,= \,1\,
+\sum_{n\,=1}^{\infty }\frac {\nu^n}{n!}\int\! ...\!\int
F_n^{(eq)}({\bf r}^{(n)}|\,{\bf R};\nu)\, \prod_{j\,=1}^n \phi({\bf
r}_j)\,d{\bf r}_j\,\,\,,\nonumber
\end{eqnarray}
where we introduced new functional argument\, $\,\phi({\bf
r})\,\equiv\,\int \psi({\bf r},{\bf p})\, G_m({\bf p})\,d{\bf
p}\,$\,\, and besides generating functional\,
$\,\mathcal{F}^{(eq)}\,$\, of equilibrium DF. It is easy to guess
that expression $\,G_M({\bf P})\,\mathcal{F}^{(eq)}\,$ should bring
stationary solution of (\ref{fe}):
\[
\left[-({\bf P}/M)\cdot\partial /\partial {\bf
R}+\mathcal{\widehat{L}}\,\right ]\,G_M({\bf P})\,
\mathcal{F}^{(eq)}\{\phi |\,{\bf R};\nu\}\,=\,0\,
\]
From here equation
\begin{equation}
\left[\frac {\partial }{\partial {\bf r}}\,+\frac
{\Phi_b^{\,\prime}({\bf r}-{\bf R})}{T}\right]\frac {\delta
\mathcal{F}^{(eq)}}{\delta \phi({\bf r})}\,=\,\frac 1T \int
[\,1+\phi({\bf r}^{\prime})\,]\,\Phi_a^{\,\prime}({\bf
r}^{\prime}-{\bf r})\, \frac {\delta^2 \mathcal{F}^{(eq)}}{\delta
\phi({\bf r})\,\delta \phi({\bf r}^{\,\prime})}\,d{\bf
r}^{\prime}\,\,\,\label{ter}
\end{equation}
follows which is analogue of equation
(2.14) in \cite{bog} and determines equilibrium DF.

Unfortunately, to the best of my knowledge, non-stationary solutions
to ``generating equations'' like (\ref{fe}) (or (7.9) from
\cite{bog}) never were investigated by Bogolyubov or any other
authors. However, past experience in the BBGKY equations (see e.g.
\cite{sil}) points to desirability of a change of variables, i.e.
transition from DF to some suitably defined ``correlation
functions''. With this purpose, let us discuss hypothetical
equalities
\[
\begin{array}{l}
F_n(t,{\bf R}, {\bf r}^{(n)},{\bf P},{\bf p}^{(n)}|\,{\bf
R}_0;\nu)\,\stackrel{?}{=}\,F_0(t,{\bf R},{\bf P}|\,{\bf R}_0;\nu)
\,\,F_n^{(eq)}({\bf r}^{(n)}|\,{\bf R};\nu)\prod_{j\,=1}^n G_m({\bf
p}_j)\,\,
\end{array}
\]
or, equivalently, \,\, $\,\mathcal{F}\{t,{\bf R},{\bf
P},\,\psi\,|\,{\bf R}_0;\nu\}\,\stackrel{?}{=}\,F_0(t,{\bf R},{\bf
P}|\,{\bf R}_0;\nu)\,\mathcal{F}^{(eq)}\{\phi\,|\,{\bf
R};\nu\}\,$\,\, (recollect that\, $\,\phi({\bf r})\,=\,\int \psi({\bf
r},{\bf p})\,G_m({\bf p})\,d{\bf p}\,$). They state that correlations
of atoms with wandering BP always stay the same as with pinned BP. It
seems reasonable in view of thermodynamically equilibrium character
of the wandering. Nevertheless, it may be true only if all possible
BP's positions are statistically equivalent. The latter in our case
is not true since translation symmetry is destroyed by information
about BP's start point $\,{\bf R}_0\,$. Correspondingly, above
equalities are incompatible with equations (\ref{fn}). For example,
substitution of equality for $\,F_1\,$ to ``collision integral'' in
equation for $\,F_0\,$ turns it into zero as if BP does not interact
with atoms at all.

The aforesaid shows that, first, BP's  wandering produces specific
non-equilibrium (in statistical sense) ``historical'' correlations
between its total path $\,\Delta{\bf R}(t)={\bf R}(t)-{\bf R}_0\,$
and current state of surrounding medium. Second, we can adequately
separate these correlations from equilibrium ones if define them as
follows:
\begin{eqnarray}
\mathcal{F}\{t,{\bf R},{\bf P},\,\psi\,|\,{\bf R}_0;\nu\}\,=
\,\mathcal{V}\{t,{\bf R},{\bf P},\,\psi\,|\,{\bf
R}_0;\nu\}\,\,\mathcal{F}^{(eq)}\{\phi\,|\,{\bf
R};\nu\}\,\,\,,\label{vf}\\
\mathcal{V}\{t,{\bf R},{\bf P},\psi\,|\,{\bf
R}_0;\nu\}\,=\,V_0\,+\sum_{n\,=1}^{\infty } \frac {\nu^n}{n!}\int_1
...\int_n V_n \,\prod_{j\,=1}^n \psi({\bf r}_j,{\bf
p}_j)\,\,\,,\nonumber
\end{eqnarray}
where\, $\,V_n=V_n(t,{\bf R}, {\bf r}^{(n)},{\bf P},{\bf
p}^{(n)}|\,{\bf R}_0;\nu)\,$
are corresponding correlation functions (CF) and \,
$\,\mathcal{V}\,$\, their GF.
In particular, evidently,\,
$\,V_0(t,{\bf R},{\bf P}| \,{\bf R}_0;\nu)\,=
\,F_0(t,{\bf R},{\bf P}|\, {\bf R}_0;\nu)\,$\, and

\begin{equation}
\begin{array}{l}
F_1(t,{\bf R},{\bf r}_1,{\bf P},{\bf p}_1|\,{\bf
R}_0;\nu)\,=\,\\
\,\,\,\,\,\,\,\,\,=\,F_0(t,{\bf R},{\bf P}|\,{\bf
R}_0;\nu)\,F_1^{(eq)}({\bf r}_1|{\bf R};\nu)\,G_m({\bf p}_1)\,+\,
V_1(t,{\bf R},{\bf r}_1,{\bf P},{\bf p}_1|\,{\bf R}_0;\nu)\,
\label{cf1}
\end{array}
\end{equation}
In terms of CF the initial conditions (\ref{ic}) and ``normalization
to volume'' conditions (``weakening of correlations'' at infinity)
take very simple form:
\begin{equation}
\begin{array}{l}
V_n(t=0)\,=\,\delta_{n,\,0}\,\,\delta({\bf R}-{\bf R}_0)
\,\,\,,\,\,\,\,\,\,\mathcal{V}\{0,\,{\bf R},{\bf P},\,\psi|\,{\bf
R}_0;\nu\}\,=\,\delta({\bf R}-{\bf R}_0)\,\,\,,\label{icv}
\end{array}
\end{equation}
\begin{equation}
\begin{array}{l}
V_{n>\,0\,}(t\,,...\,\,{\bf r}_k\rightarrow
\infty\,\,...\,)\,\rightarrow \,0\,\,\label{bcv}
\end{array}
\end{equation}
Substituting (\ref{vf}) into (\ref{fe}) one obtains
equation for GF of  ``historical'' correlations:
\begin{equation}
\frac {\partial \mathcal{V}}{\partial t}\,+\,\frac {\bf
P}{M}\cdot\frac {\partial \mathcal{V}}{\partial {\bf
R}}\,=\,\widehat{\mathcal{L}}\left(\psi,\frac {\delta
}{\delta\psi}\right)\,\mathcal{V}\,+
\,\widehat{\mathcal{L}}^{\,\,\prime}\left(\nu,\psi,\frac {\delta
}{\delta\psi}\right)\,\mathcal{V}\,\,\,,\,\label{fev}
\end{equation}
\begin{eqnarray}
\widehat{\mathcal{L}}^{\,\,\prime}\left(\nu,\psi,\frac {\delta
}{\delta\psi}\right)\,=\,\left\{\int [\,1\,+\phi({\bf
r})\,]\,\,\Phi_b^{\,\prime}({\bf R}-{\bf
r})\,\,\nu\,\mathcal{C}\{{\bf r},\phi\,|\,{\bf R};\,\nu\}\,\,d{\bf
r}\right\}\left(\frac {{\bf P}}{MT}+\frac {\partial
}{\partial {\bf P}}\right )\,+\,\,\,\,\nonumber\\
+\,\int_1\int_2\,\,
[\,1+\psi(x_1)\,]\,[\,1+\psi(x_2)\,]\left[\,\Phi_a({\bf r}_1-{\bf
r}_2)\,,\,\nu\,\mathcal{C}\{{\bf r}_2,\phi\,|\,{\bf
R};\,\nu\}\,G_m({\bf p}_2)\,\frac {\delta }{\delta \psi(x_1)}
\,\right]\nonumber\,\,\,,
\end{eqnarray}
where new functional\,
$\,\mathcal{C}\,$\, is defined as
\begin{equation}
\mathcal{C}\{{\bf r},\phi\,|\,{\bf R};\,\nu\}\,=\,\frac {\delta \ln
\mathcal{F}^{(eq)}\{\phi |\,{\bf R};\,\nu\}}{\nu\,\delta \phi({\bf
r})}\,=\,\,\,\,\,\,\,\,\,\,\,\,\,\,\,\,\,\,\,\,\,\,\,\,\label{c}
\end{equation}
\[
=\,F_1^{(eq)}({\bf r}|\,{\bf R};\nu)+\sum_{n\,=\,1}^{\infty }\frac
{\nu^n}{n!}\int\!\! ...\!\!\int C_{n+1}({\bf r},{\bf r}_1...\,{\bf
r}_n|\,{\bf R};\nu) \prod_{j\,=1}^n \phi({\bf r}_j)\,d{\bf r}_j\,\,
\]
Complication of equation (\ref{fev}) in comparison with (\ref{fe}) is
pay for simple conditions (\ref{icv})-(\ref{bcv}). Corresponding
equations for CF also are more complicated than (\ref{fn}). Therefore
here we write out them only for extreme but interesting case of ``BP
in ideal gas''  (when $\,\Phi_a({\bf r})=0\,$, i.e. atoms do not
interact with themselves):
\begin{eqnarray}
\frac {\partial V_0}{\partial t}\,=\,-\frac {\bf P}{M}\cdot\frac
{\partial V_0}{\partial {\bf R}}\,+\,\nu\, \frac {\partial }{\partial
{\bf P}}\int_{1} \Phi^{\prime}_b({{\bf R}-\bf
r}_{1})\,V_{1}\,\,\,,\nonumber
\end{eqnarray}
\begin{eqnarray}
\frac {\partial V_{n>\,0}}{\partial t}\,=\,[\,H_n\,,V_n\,]\,+\,\nu\,
\frac {\partial }{\partial {\bf P}}\int_{n+1} \Phi^{\prime}_b({{\bf
R}-\bf r}_{n+1})\,V_{n+1}\,\,+\,\,\,\,\,\,\,\,\,\,\,\,
\,\,\,\,\,\,\label{vn}
\end{eqnarray}
\begin{eqnarray}
\,\,\,\,\,\,\,\,\,\,\,\,\,\,\,\,\,+\,\,T\,\sum_{j\,=1}^{n}\,
\mathcal{P}(j,n)\,\,G_m({\bf p}_n)\,E^{\,\prime}({\bf r}_n-{\bf
R})\left(\frac {{\bf P}}{MT}+\frac {\partial }{\partial {\bf
P}}\right ) V_{n-\,1}\,\,\,,\nonumber
\end{eqnarray}
where\, $\,H_n ={\bf P}^2/2M + \sum_{j\,=1}^n\,[\,{\bf p}_j^2/2m
+\Phi_b({{\bf R}-\bf r}_j)\,]\,$\,,\,  $\,E({\bf
r})=\exp{[\,-\,\Phi({\bf r})/T\,]}\,$,\, $\,E^{\,\prime}({\bf
r})=\nabla E({\bf r})=-\,\Phi^{\,\prime }({\bf r})E({\bf r})/T\,$\,
and\, $\,\mathcal{P}(j,n)\,$\, denotes operation
of transposition of arguments\,
$\,x_j\,=\,\{{\bf r}_j,{\bf p}_j\}\,$\, and
$\,x_n\,=\,\{{\bf r}_n,{\bf p}_n\}\,$.\,
Thus  natively bidiagonal BBGKY hierarchy in
terms of CF becomes tridiagonal.

\section{Invariance group of equilibrium generating functional}
Equilibrium CF\, $\,C_{n}\,$ in (\ref{c}) are determined by first
equality in (\ref{c}):
\[
\begin{array}{l}
C_2({\bf r},{\bf r}_1|\,{\bf R};\nu)\,= \,F_2^{(eq)}({\bf r},{\bf
r}_1|{\bf R};\nu)-F_1^{(eq)}({\bf r}|{\bf
R};\nu)\,F_1^{(eq)}({\bf r}_1|{\bf R};\nu)\,\,\,,\\
C_3({\bf r},{\bf r}_1,{\bf r}_2)\,=\,F^{(eq)}_3({\bf r},{\bf
r}_1,{\bf r}_2)\,+\,2\,F^{(eq)}_1({\bf r})\,F^{(eq)}_1({\bf
r}_1)\,F^{(eq)}_1({\bf r}_2)\,-\\
\,\,\,\,\,\,\,\,\,\,\,\,\,\,\,\, \,\,\,\,\,\,\,-\,F^{(eq)}_1({\bf
r})\,F^{(eq)}_2({\bf r}_1,{\bf r}_2)\,-\,F^{(eq)}_1({\bf
r}_1)\,F^{(eq)}_2({\bf r},{\bf r}_2)\,-\,F^{(eq)}_1({\bf
r}_2)\,F^{(eq)}_2({\bf r},{\bf r}_1)\,\,\,,
\end{array}
\]
and so on. In the second of these expressions for brevity we omitted
arguments $\,{\bf R}\,$ and $\,\nu\,$. The conditions of
normalization to volume, i.e. weakening of correlations at infinity,
which in essence establish existence of the thermodynamical limit
\cite{bog,rue}, do mean  that all these CF vanish when distance
between any two atoms goes to infinity and turn into usual
equilibrium CF when BP is moved off to infinity:
\begin{equation}
\begin{array}{l}
C_{n+1}({\bf r},{\bf r}_1\,...\,{\bf r}_n|\,{\bf
R};\nu)\,\rightarrow\,0\,\,\,\,\,\,\,\,\,\texttt{at}
 \,\,\,\,\,\,\,\,{\bf r}_j-{\bf r}
 \rightarrow\infty \,\,\,,\label{cbc}\\
 C_{n}({\bf r}_1\,...\,{\bf r}_n|\,{\bf
R};\nu)\,\rightarrow\,C_{n}({\bf r}_1\,...\,{\bf
r}_n;\,\nu)\,\,\,\,\,\,\,\,\texttt{at} \,\,\,\,\,\,\,\,\,{\bf R}-{\bf
r}_j\rightarrow\infty\,\,
\end{array}
\end{equation}
Moreover, under sufficiently short-range interactions all these
limits are achieved in a fast (absolutely) integrable way, which will
be assumed below.

Next, consider equation  (\ref{ter}) rewriting it in the form
\[
\left[\frac {\partial }{\partial {\bf r}}+\frac
{\Phi_b^{\,\prime}({\bf r}-{\bf R})}{T}\right]\mathcal{C}\{{\bf
r},\phi\,|{\bf R};\,\nu\}\,=\,\frac 1T \int [\,1+\phi({\bf
r}^{\prime})\,]\, \Phi_a^{\,\prime}({\bf r}^{\prime}-{\bf r})\,\frac
{\delta \mathcal{C}\{{\bf r}^{\,\prime},\phi\,|{\bf
R};\,\nu\}}{\delta \phi({\bf r})}\,\,d{\bf r}^{\prime}\,+\nonumber
\]
\begin{eqnarray}
+\,\,\mathcal{C}\{{\bf r},\phi\,|{\bf R};\,\nu\}\,\,\frac
{\nu}{T}\int [\,1+\phi({\bf r}^{\prime})\,]\,\Phi_a^{\,\prime}({\bf
r}^{\prime}-{\bf r})\,\mathcal{C}\{{\bf r}^{\prime},\phi\,|{\bf
R};\,\nu\}\,d{\bf r}^{\,\prime}\,\,\,\,\,\,\,\label{ter1}
\end{eqnarray}
as equation for the functional\, $\,\mathcal{C}\,$. The latter
interests us here only so far as it influences equation (\ref{fev})
for GF of ``historical correlations''. In this respect, we have
(perhaps, for the first time) to pay attention to some important
properties of this GF. Firstly, due to integrability of the
asymptotic (\ref{cbc}), $\,\mathcal{C}\,$ can be extended to bounded
functions $\,\phi({\bf r})\,$ which do not turn to zero at infinity,
particularly, to constants, and introduce objects as follow:
\begin{equation}
C(\sigma ,\nu)=\lim_{{\bf R}-\,{\bf r}\,\rightarrow\infty}
\,\mathcal{C}\{{\bf r},\sigma|{\bf R};\nu\}
=1+\sum_{n\,=\,1}^{\infty
}\frac {\nu^n\sigma^n}{n!}\int_1\! ...\!\int_n
C_{n+1}({\bf r},{\bf
r}_1\,...\,{\bf r}_n;\nu)\,\,\,,\label{ccc}
\end{equation}
\begin{equation}
\mathcal{C}_{\sigma}\{{\bf r},\phi\,|{\bf R};\nu\}\,=\,\frac
{\mathcal{C}\{{\bf r},\sigma +\phi\,|{\bf R};\,\nu\}}{C(\sigma
,\nu)}\,\,\,,\label{nc}
\end{equation}
where  $\,\sigma=\,$const\,, $\,\int_n...\,=\,\int...\,\,d{\bf
r}_n\,$\,, and integrals in (\ref{ccc}) are factually
independent on $\,\,{\bf r}\,$.
Secondly, replacement
$\,\phi({\bf r})\rightarrow\sigma +\phi({\bf r})\,$
in (\ref{ter1}) and elementary algebraic manipulations
transform (\ref{ter1}) to equation
for $\,\mathcal{C}_{\sigma}\,$\,:
\[
\left[\frac {\partial }{\partial {\bf r}}+\frac
{\Phi_b^{\,\prime}({\bf r}-{\bf
R})}{T}\right]\mathcal{C}_{\sigma}\{{\bf r},\phi\,|{\bf
R};\nu\}=\frac 1T\! \int \!\!\left[1+\frac {\phi({\bf
r}^{\prime})}{1+\sigma}\right] \Phi_a^{\,\prime}({\bf r}^{\prime}-{\bf
r})\,\frac {\delta\, \mathcal{C}_{\sigma}\{{\bf
r}^{\,\prime},\phi\,|{\bf R};\nu\}}{\delta\,[\, \phi({\bf
r})/(1+\sigma )\,]}\,\,d{\bf r}^{\prime}\,+
\]
\begin{equation}
+\,\mathcal{C}_{\sigma}\{{\bf r},\phi\,| {\bf R};\nu\}\,\,\frac
{\nu\,C(\sigma ,\nu)\,(1+\sigma)}{T}\int\!  \left[1+\frac {\phi({\bf
r}^{\prime})}{1+\sigma}\right] \Phi_a^{\,\prime}({\bf r}^{\prime}-{\bf
r})\,\,\mathcal{C}_{\sigma}\{{\bf r}^{\,\prime},\phi\,|{\bf
R};\nu\}\,\,d{\bf r}^{\,\prime}\,\,\label{ter2}
\end{equation}
It differs from (\ref{ter1}) only by  scale transformation
of the functional argument,\,
$\,\phi({\bf r})\rightarrow \phi({\bf r})/(1+\sigma )\,$\,,
and replacement of the density\, $\,\nu\,$\, by
\begin{equation}
\begin{array}{l}
\upsilon (\sigma ,\nu)\,=\,\nu\,C(\sigma
,\nu)\,(1+\sigma)\,\,\label{nd}
\end{array}
\end{equation}
Third, formal solution to (\ref{ter1}) in the form of series
(\ref{c}) is unambiguously determined by the normalization
conditions, that is conditions (\ref{cbc}) plus quite obvious
equality\, $\,\lim_{\,\,{\bf R}-\,{\bf r}\,\rightarrow\,\infty}\,
\mathcal{C}\{{\bf r},\phi =0\,|{\bf R};\,\nu\}=1\,$\, which
determines first term of the series. Fourth, functional
$\,\mathcal{C}_{\sigma}\,$ defined by (\ref{ccc})-(\ref{nc})
satisfies same conditions since (\ref{cbc}) imply
\begin{equation}
\sum_{k\,=0}^{\infty}\frac {\nu^k\sigma^k}{k!} \int_{n+1}\!
...\!\int_{n+k} C_{n+k+1}({\bf r},{\bf r}_1\,...\,{\bf
r}_{n+k}|\,{\bf R};\nu)\,\rightarrow\,0\,
\,\,\,\,\,\texttt{at}\,\,\,\,\,{\bf r}_j-{\bf
r}\rightarrow\infty\,\,\,,\label{cbc1}
\end{equation}
at least, if one understands  (\ref{cbc}) in the sense of absolute
integrability of CF $\,C_{n+1}\,$ and speaks about not too large
values of the density \footnote{\,An useful information in this
respect comes from rigorous consideration of ``group properties of
correlation functions'' \cite{rue}.}.

Summarizing all that, we can conclude that solution to equation
(\ref{ter2}) is nothing but\,\, $\,\mathcal{C}_{\sigma}\{{\bf
r},\phi\,|{\bf R},\nu \}\,= \,\mathcal{C}\{{\bf r},\phi/(1+\sigma
)\,|\,{\bf R};\,\upsilon (\sigma ,\nu) \}\,$\,.\, With taking into
account (\ref{nc}) and (\ref{nd}) this means that at arbitrary
(admissible in definite sense) constant $\,\sigma $ and bounded
function $\,\phi=\phi({\bf r})\,$ equality
\begin{equation}
\nu\,\mathcal{C}\{{\bf r},\sigma +\phi\,|\,{\bf R};\,\nu\}\,=\,\frac
{\upsilon (\sigma ,\nu)}{1+\sigma}\,
\,\mathcal{C}\left\{{\bf r},\frac
{\phi}{1+\sigma}\,|\,{\bf R};\,\upsilon (\sigma
,\nu)\right\}\,\,\label{eqid}
\end{equation}
is valid\,. It can be rewritten also as
\begin{equation}
\widehat{\mathcal{T}}(\sigma)\,\mathcal{C}\{{\bf r},\phi\,|\,{\bf
R};\,\nu\}\,\equiv\,C(\sigma ,\nu )\,\,\mathcal{C}\left\{{\bf
r},\frac {1+\phi}{1+\sigma}-1\,|\,{\bf R};\,\upsilon(\sigma
,\nu)\right\}\,= \,\mathcal{C}\{{\bf r},\phi\,|\,{\bf
R};\,\nu\}\,\,\,,\label{gr}
\end{equation}
where functions $\,C(\sigma ,\nu )\,$ and $\,\upsilon(\sigma ,\nu
)\,$ are connected by means of (\ref{nd}), and the left equality
defines one-parameter family of such transformations of arguments of
functional $\,\,\mathcal{C}\{{\bf r},\phi\,|\,{\bf R};\,\nu\}\,$
which do not change its value. It is not hard to verify that this is
group described by the composition rules
\begin{eqnarray}
\widehat{\mathcal{T}}(\sigma_2)\,\widehat{\mathcal{T}}(\sigma_1)\,=
\,\widehat{\mathcal{T}}(\sigma_1 +\sigma_2 +\sigma_1\sigma_2)
\,\,\,,\nonumber\\ \upsilon (\sigma_2\,,\,\upsilon (\sigma_1\,
,\nu))\,=\,\upsilon
(\sigma_1 +\sigma_2 +\sigma_1\sigma_2 \,,\nu)\,\,\,,\label{gr1}\\
C(\sigma_2\,,\,(1+\sigma_1)\,C(\sigma_1\, ,\nu)\,\nu)\,C(\sigma_1\,
,\nu)\,=\,C(\sigma_1 +\sigma_2 +\sigma_1\sigma_2 \,,\nu)\,\nonumber
\end{eqnarray}
with restrictions\,  $\,\sigma >-1\,$\,,\, $\,\phi({\bf r}) >-1\,$\,.
The latter are clear in the light of that
\[
\upsilon\{{\bf r}|\,\phi ,{\bf R};\nu \} \equiv [1+\phi({\bf
r})]\,\frac {\delta \ln\mathcal{F}^{(eq)}\{\phi |\,{\bf
R};\nu\}}{\delta\phi({\bf r})}\,=\,\nu\,[1+\phi({\bf
r})]\,\mathcal{C}\{{\bf r},\phi\,|\,{\bf R};\,\nu\}\,
\]
represents mean concentration of atoms under external potential
$\,U({\bf r})\,$ which is related to $\,\phi({\bf r})\,$ by
$\,\phi({\bf r})=\exp{[\,-\,U({\bf r})/T\,]}-1\,$\, (see e.g.
\cite{mpa3}). Substitution $\,\sigma =\exp{(a)}-1\,$ gives
$\,\widehat{\mathcal{T}}(a_2)\, \widehat{\mathcal{T}}(a_1)=
\widehat{\mathcal{T}}(a_1 +a_2)\,$ thus eliminating the restrictions.

Infinitesimal form of  (\ref{eqid}) or (\ref{gr}) looks best if
written through particular CF\,:
\begin{eqnarray}
\left\{\varkappa(\nu)+[\,1+\varkappa(\nu)\,] \,\nu\,\frac
{\partial}{\partial \nu}\right\}F_1^{(eq)}({\bf r}|\,{\bf
R};\nu)\,=\,\nu\!\int\! C_2({\bf r},{\bf r}^{\,\prime}|{\bf
R};\nu)\,d{\bf r}^{\,\prime}\,\,\,,
\,\,\,\,\,\,\,\,\,\,\,\,\,\nonumber\\
\left\{n\varkappa(\nu)+[\,1+\varkappa(\nu)\,] \,\nu\,\frac
{\partial}{\partial \nu}\right\} C_{n}({\bf r}_1\,...\,{\bf
r}_n|\,{\bf R};\nu)=\nu\!\int\! C_{n\,+1}({\bf r}_1...\,{\bf
r}_{n},{\bf r}^{\,\prime}|{\bf R};\nu)\,d{\bf r}^{\,\prime}
\,\,\,,\nonumber\\
\varkappa(\nu)\,\equiv \,\left[\frac {\partial C(\sigma
,\nu)}{\partial \sigma}\right]_{\sigma =0}\,=\,\nu\int C_2({\bf
r},0;\,\nu)\,d{\bf r}\,\,\,\,\,\,\,\,\,\,\,\,\,\,
\,\,\,\,\,\,\,\,\,\, \,\,\,\,\,\,\,\,\,\, \,\,\,\,\, \label{eqinf}
\end{eqnarray}
The function\, $\,\varkappa(\nu)\,$\, is known (see e.g. \cite{ll1})
to be directly related to a state equation of the system:\,
$\,1+\varkappa(\nu)\,=\,T\,(\partial \nu/\partial
\mathcal{P})_T\,$\,, where\, $\,\mathcal{P}\,$ denotes the pressure.
Notice that in the framework of the grand canonical ensemble
substantially similar relations can be easy derived by
differentiation of DF in respect to the activity.

\section{Invariance group of generating functional \\
of historical correlations} Now, let us show that solution to
equation (\ref{fev}) has invariance properties quite similar to
(\ref{gr}). Since initial condition to this equation (see
(\ref{icv})) does not depend on variables $\,\psi =\psi({\bf r},{\bf
p})\,$ and $\,\nu\,$\, at all, solution to it is completely
determined by structure of operators $\,\widehat{\mathcal{L}}\,$ and
$\,\widehat{\mathcal{L}}^{\,\,\prime}\,$ and conditions (\ref{bcv}).
The latter formally allow to extend functional $\,\mathcal{V}\{t,{\bf
R},{\bf P},\psi\,|{\bf R}_0;\nu\}\,$ (like $\,\mathcal{C}\,$ before)
to arguments  $\,\sigma +\psi({\bf r},{\bf p})\,$, with $\,\sigma
=\,$\,const\,, in place of $\,\psi({\bf r},{\bf p})\,$. The fact that
limit in (\ref{bcv}) is achieved fast enough to indeed ensure this
extension can be confirmed afterwards. Besides, thanks to (\ref{bcv})
variable $\,\psi(x_1)\,$ in expression\,
$\,\widehat{\mathcal{L}}\,\mathcal{V}\,$\, inside (\ref{fev}) (see
definition (\ref{l}) of the operator $\,\widehat{\mathcal{L}}\,$) can
be shifted by arbitrary constant:
\[
\int_1 \psi(x_1)\,\, \frac {{\bf p}_1}{m}\cdot\frac {\partial
}{\partial {\bf r}_1}\,\frac {\delta \mathcal{V}}{\delta
\psi(x_1)}\,\,=\,\int_1 \,[\,a +\psi(x_1)\,]\,\, \frac {{\bf
p}_1}{m}\cdot\frac {\partial }{\partial {\bf r}_1}\,\frac {\delta
\mathcal{V}}{\delta \psi(x_1)}\,\,\,,
\]
where\, $\,a =\,$\,const\,, for instance,\, $\,a =1\,$\,.
This is important difference of\,
$\,\widehat{\mathcal{L}}\,\mathcal{V}\,$\,
from $\,\mathcal{\widehat{L}}\,\mathcal{F}\,$.
Consequently, taking in mind action of
$\,\widehat{\mathcal{L}}\,$ onto $\,\mathcal{V}\,$,
one can write
\begin{equation}
\mathcal{\widehat{L}}\left(\sigma +\psi\,,\frac {\delta }{\delta
\psi}\right )\,=\,\mathcal{\widehat{L}}\left(\frac
{\psi}{1+\sigma}\,,\frac {\delta }{\delta\, [\,\psi/(1+\sigma
)\,]}\right )\,\,\label{lt}
\end{equation}
Further, let us carefully consider operator
$\,\mathcal{\widehat{L}}^{\,\,\prime}\,$ (see also (\ref{fev})). In
contrast to $\,\mathcal{\widehat{L}}\,$, it depends on the density
$\,\nu\,$. Nevertheless, with the help of equality (\ref{eqid}) it is
easy to make sure that it obeys the same relation if transformation
of argument $\,\psi({\bf r},{\bf p})\,$ is accompanied by
transformation of argument $\,\nu\,$ in accordance with (\ref{nd})
and (\ref{ccc}):
\begin{equation}
\mathcal{\widehat{L}}^{\,\,\prime}\left(\nu\,,
\,\sigma +\psi\,,\frac
{\delta }{\delta \psi}\right
)\,=\,\mathcal{\widehat{L}}^{\,\,\prime}
\left(\,\upsilon(\sigma,\nu)\,,\,\frac
{\psi}{1+\sigma}\,,\frac {\delta }{\delta\, [\,\psi/(1+\sigma
)\,]}\right )\,\,\label{lpt}
\end{equation}
Formulas (\ref{lt}) and (\ref{lpt}) just imply
the noted invariance property of solutions of  (\ref{fev}):
\begin{equation}
\mathcal{V}\{t,{\bf R},{\bf P},\,\sigma +\psi\,|\,{\bf
R}_0;\,\nu\}\,=\,\mathcal{V}\left\{t,{\bf R},{\bf P},\,\frac
{\psi}{1+\sigma}\,|\,{\bf R}_0;\,\upsilon (\sigma
,\nu)\right\}\,\,\,\label{id}
\end{equation}
or, equivalently and similarly to (\ref{gr}),
\begin{eqnarray}
\widehat{\mathcal{T}}(\sigma)\,\mathcal{V}\{t,{\bf R},{\bf
P},\,\psi\,|\,{\bf R}_0;\,\nu\}\,\equiv\,\mathcal{V}\{t,{\bf R},{\bf
P},\,\frac {1+\psi}{1+\sigma}-1\,|\,{\bf R}_0;\,\upsilon (\sigma
,\nu)\}\,=\,\nonumber\\
=\,\mathcal{V}\{t,{\bf R},{\bf P},\,\psi\,|\,{\bf
R}_0;\,\nu\}\,\,\,,\,\,\,\,\,\,\,\,\, \,\,\,\, \,\,\,\, \,\,\,\,
\,\,\,\, \,\,\,\, \,\,\,\, \,\,\,\, \,\,\,\,\label{grv}
\end{eqnarray}
where left inequality together with
(\ref{nd}) and (\ref{ccc}) defines action of
the above described group onto GF
of historical correlations.
Expansion of (\ref{id}) into series
over $\,\psi\,$ yields
\begin{eqnarray}
V_0(t,{\bf R},{\bf P}|\,{\bf R}_0;\,\upsilon (\sigma ,\nu))\,
=\,V_0(t,{\bf R},{\bf P}|\,{\bf R}_0;\,\nu)\,+\,\,\,\,\,\,\,
\,\,\,\,\,\,\,\,\,\,\,\,\, \,\,\,\,\,\,\,\,\,\,\,\,\,\,\,\,\,
\,\,\,\,\,\,\,\,\,\,\,\,\, \,\,\,\,\,\,\,\,\,\,\,\, \,\,\, \label{vexp}\\
\,\,\,\,\,\,\,\,\,\,\,\, \,\,\,\,\,\,\,\,\,\,\,\,\,\,\,\,\,\,\,\,
\,\,\,\,\,\,\,\,\, \,\,\, \,\,\,\,\,\,\,\, \,\,\,\,\,\,\,\,\
\,\,\,\,\,\,\,\,\,\,\,\,\ +\,\sum_{n\,=\,1}^\infty \frac
{\nu^n\sigma^n}{n!}\,\int_1\! ...\!\int_n V_n(t,{\bf R},{\bf
r}^{(n)},{\bf P},{\bf p}^{(n)}|\,{\bf R}_0;\,\nu)\,\,\,,\nonumber\\
\left[\frac {\upsilon (\sigma ,\nu)}
{(1+\sigma)\nu}\right]^k\,V_k(t,{\bf R},{\bf r}^{(k)},{\bf P},{\bf
p}^{(k)}|{\bf R}_0; \,\upsilon (\sigma ,\nu))\,=\, V_k(t,{\bf R},{\bf
r}^{(k)},{\bf P},{\bf p}^{(k)}|{\bf
R}_0;\,\nu)\,+\nonumber\\
+\sum_{n\,=\,1}^\infty \frac {\nu^n\sigma^n}{n!}\int_{k+1}\! ...
\!\int_{k+n} V_{k+n}(t,{\bf R}, {\bf r}^{(k+n)},{\bf P},{\bf
p}^{(k+n)}|{\bf R}_0;\,\nu)\,\,\label{vexpn}
\end{eqnarray}
Corresponding infinitesimal (in respect to $\,\sigma\,$)
relations are similar to (\ref{eqinf})\,:
\begin{eqnarray}
\left\{n\,\varkappa(\nu)\,+\,[\,1+\varkappa(\nu)\,] \,\nu\,\frac
{\partial}{\partial \nu}\,\right\}\,V_n(t,{\bf R},{\bf r}^{(n)},{\bf
P},{\bf p}^{(n)}|{\bf R}_0;\,\nu)\,=\,\,\,
\,\,\,\,\,\,\,\,\,\,\,\,\,\,\,\,\,\,\label{inf}\\
\,\,\,\,\,\,\,\,\,\,\,\, \,\,\,\,\,\,\,\,\,\,\,\,\,\,\,\,\,\,\,
\,=\,\nu\int_{n+1} V_{n+1}(t,{\bf R},{\bf r}^{(n+1)},{\bf P},{\bf
p}^{(n+1)}|{\bf R}_0;\,\nu) \,\nonumber
\end{eqnarray}

Formulas (\ref{vexp})-(\ref{vexpn}) can be interpreted as ``virial
expansions'' of probabilistic law of BP's random wandering and
historical correlations between BP and medium, with those difference
from usual virial expansions of thermodynamic quantities  \cite{ll1}
or kinetic coefficients \cite{ll2} that here decrements of the
density do figure instead of its full value. However, in the limit
$\,\nu\rightarrow 0\,$, $\,\sigma\rightarrow \infty\,$, $\,\nu\sigma
=\,$const\, these relations take quite usual form. The simplest of
them, (\ref{vexp}), recently was obtained \cite{mpa3,pro,jstat} in
other way, starting from the ``generalized fluctuation-dissipation
relations''  \cite{fds,p}.

As far as I know, exact relations of this kind never before were
under consideration.  It would be rather hard to extract equalities
(\ref{inf}) (all the more, (\ref{vexp})-(\ref{vexpn})) directly from
the BBGKY equations (\ref{fn}). This is possible only in the case of
``BP in ideal gas'' (the reader can try to see how (\ref{inf})
follows directly from equations (\ref{vn}) for CF).

\vspace{-10 pt}
\section{On principal consequences from the virial relations}
All the exact ``virial relations'' (\ref{vexp})-(\ref{inf}) are
automatically satisfied by exact solution to BBGKY equations.
Therefore these relations can be applied to ``testing of statistical
hypotheses'' about the solution or constructing approximations to it.
Moreover, in the just mentioned limit $\,\nu\rightarrow 0\,$,
$\,\sigma\rightarrow \infty\,$, $\,\nu\sigma
=\nu^{\,\prime}=\,$const\,, with the help of formulas  (\ref{ccc}),
(\ref{nc}) and (\ref{nd}), it is easy to transform equalities
(\ref{vexp})-(\ref{vexpn}) into explicit formal representation of
exact solution of BBGKY hierarchy for $\,V_n(t,{\bf R},{\bf
r}^{(n)},{\bf P},{\bf p}^{(n)}|{\bf R}_0;\,\nu^{\,\prime}\,)\,$ as a
power series in respect to $\,\nu^{\,\prime}\,$.

It should be emphasized that (\ref{vexp})-(\ref{vexpn}) connect
random walks of BP in two media whose densities may form arbitrary
large or arbitrary small ratio, $\,\upsilon (\sigma ,\nu)/\nu\,$.
Therefore, any cutting off the series in (\ref{vexp})-(\ref{vexpn})
would present not an approximate but incorrect result. This means
that from the viewpoint of exact theory all historical correlations
always are equally significant. Even in the ``Boltzmann-Grad limit''
(``dilute gas limit'') when gas parameters tend to zero, $\,4\pi
r_a^3\nu/3\rightarrow 0\,$, $\,4\pi r_b^3\nu/3\rightarrow 0\,$
(with\, $\,r_{a,\,b}\,$\, standing for radii of interactions of atoms
and BP) while free paths $\,\Lambda_b= (\pi r_b^2\nu)^{-1}\,$ and
$\,\Lambda_a= (\pi r_a^2\nu)^{-1}\,$\, stay fixed, or in the limit of
ideal gas (where
 $\,\Lambda_a=\infty\,$ but
$\,\Lambda_b\,$ is finite).
In both cases, formulas (\ref{vexp})-(\ref{inf})
simplify to
\begin{eqnarray}
\frac {\partial^{\,k} V_n(t,{\bf R},{\bf r}^{(n)},{\bf P},{\bf
p}^{(n)}|\,{\bf R}_0;\,\nu)}{\partial\nu^{\,k}}\, =\,\int_{n+1}\! ...
\!\int_{n+k} V_{n+k}(t,{\bf R}, {\bf r}^{(n+k)},{\bf P},{\bf
p}^{(k+n)}|{\bf R}_0;\,\nu)\,\,\label{lm}
\end{eqnarray}
We must conclude that Boltzmann's kinetics which does not know
correlations (except may be two-particle one) is not true
``zero-order approximation'' in respect do density.

As an illustration, let us consider BP in ideal gas and apply
relations (\ref{lm}) to test ``statistical hypothesis'' that a
correct approximate solution of exact equations  (\ref{vn}) can be
obtained under neglect of three-particle and higher correlations.
Such hypothesis always is (directly or indirectly) involved into
derivation of kinetic equations for ``sufficiently rarefied'' gas.
Examples can be found e.g. in \cite{bog,sil,ll2,re,vblls}\,
\footnote{\, The work \cite{vblls} on gas of hard spheres gave
example of illusory discard of correlations, as it is discussed in
\cite{last}.}. Setting $\,V_2=0\,$ in the second ($\,n=1\,$) of
equations (\ref{vn}) and then inserting result of its integration to
the first one ($\,n=0\,$) we come to closed Boltzmann-Lorentz kinetic
equation \cite{re,vblls} for $\,V_0(t,{\bf R},{\bf P}|\,{\bf
R}_0;\,\nu)\,$\,. Asymptotic of solution to this equation at $\,t\gg
\tau =\Lambda_b /v_0 \,$\, (with $\,v_0\sim\sqrt{T/M}\,$) is
wittingly Gaussian:
\begin{equation}
V_0(t,\Delta{\bf R};\nu)\equiv\!\int V_0(t,{\bf R},{\bf P}|{\bf
R}_0;\nu)\,d{\bf P}\,\rightarrow\,\frac {\exp{(-\Delta {\bf
R}^2/4Dt)}}{(4\pi Dt)^{\,3/2}}\,\,\,, \label{vg}
\end{equation}
where $\,D\,$ is BP's diffusivity,\, $\,D=v_0\Lambda_b\propto
\nu^{-1}\,$\,. Undoubtedly, this is a complicated function of the
density\, $\,\nu\,$\,. At the same time, in the view of exact
relations (\ref{lm}) the statement $\,V_2=0\,$ implies that
$\,V_0(t,\Delta{\bf R};\,\nu)\,$ should be purely linear function
of\, $\,\nu\,$\, !

So strong discrepancy prompts that our hypothesis is erroneous,
asymptotic  (\ref{vg}) is doubtful, and thus we should return to
BBGKY equations. More correct approach to their approximate solving,
with including correlations of any order, was suggested in \cite{i1}
(or see \cite{i2}) and developed in \cite{p1}. It confirmed the guess
\cite{bk12} that actual molecular random walk represents a diffusive
random process possessing scaleless ``flicker'' (i.e. with $\,1/f$
-type spectrum \cite{bk3}) fluctuations of diffusivity (as well as
mobility) of BP. In corresponding asymptotic of $\,V_0(t,\Delta{\bf
R};\,\nu)\,$, in contrary to (\ref{vg}), the exponential is replaced
by a function with power-law long tails (cut off at distances $\,\sim
v_0t\,$\,) \cite{p1}\,:
\begin{equation}
V_0(t,\Delta {\bf R};\,\nu)\,\rightarrow \,\,\frac {\Gamma
(7/2)}{[4\pi D\, t\,]^{\,3/2}}\, \left[\,1+\frac {\Delta{\bf
R}^2}{4D\,t}\,\right]^{-\,7/2}\Theta \left (\frac {|\Delta {\bf
R}|}{v_0 t}\right )\,\,\,,\label{as}
\end{equation}
where\, $\,\Theta(x)\approx 1\,$ at $\,x< 1\,$ and\,
$\,\Theta(x)\rightarrow 0\,$ at $\,x>1\,$ in a very fast way.

Origin of the diffusivity fluctuations is trivial. Heuristically,
that is indifference of the system to a number and relative frequency
of BP's collisions with atoms (all the more, to distribution of
collisions over impact parameter values) \cite{i1,i2}. At that,
higher historical correlations described by $\,V_{n>\,1}\,$ are
caused by complicity of particles in (uncontrolled and therefore
scaleless and may be non-ergodic) fluctuations of relative frequency
of BP's collisions (see also \cite{mpa3,jstat,last,p1}).

Let us demonstrate that one can come to analogous conclusions without
solving BBGKY equations but instead resting upon their invariance
group and besides the trivial fact that all the DF $\,F_n\,$ always
are non-negative.

From non-negativity of $\,F_1\,$ and identity (\ref{cf1}) we have
\begin{eqnarray}
V_0(t,\Delta {\bf R};\,\nu)\int_{\Omega} F_1^{(eq)}({\bf r}|{\bf
R};\nu)\,d{\bf r}\,+\int_{\Omega}\,V_1(t,{\bf R},{\bf r}|{\bf
R}_0;\,\nu)\,d{\bf r} \,\geq\,0\,\,\,,\label{in0}
\end{eqnarray}
where\,\, $\,V_1(t,{\bf R},{\bf r}|{\bf R}_0;\,\nu)\, \equiv\,\int
\int\, V_1(t,{\bf R},{\bf r},{\bf P},{\bf p}|{\bf R}_0;\,\nu)\,\,
d{\bf p}\,d{\bf P}\,$\,\,, and\, $\,\Omega\,$ is any region in the
space of vectors\,  $\,{\bf r}-{\bf R}\,$\,.  Introduce
$\,\Omega(\delta,t,\Delta{\bf R};\nu)\equiv \Omega(\delta)\,$ to be
minimum (in the sense of volume) of all regions $\,\Omega\,$  what
satisfy condition
\begin{eqnarray}
\left|\int_{\Omega} V_1\,d{\bf r}\,-\,\int V_1\,d{\bf
r}\,\right|\,\leq\,\delta\,\left|\int V_1\,d{\bf
r}\,\right|\,\,\,,\label{df}
\end{eqnarray}
where $\,0<\delta <1\,$. One can easy justify that (\ref{in0}) and
(\ref{df}) together imply inequality
\begin{equation}
V_0(t,\Delta {\bf R};\,\nu)\int_{\Omega(\delta)} F_1^{(eq)}({\bf
r}|{\bf R};\nu)\,d{\bf r}\,+\,(1-\delta)\int V_1(t,{\bf R},{\bf
r}|{\bf R}_0;\,\nu)\,d{\bf r}\,\,\geq\,\,0\,\,\label{in}
\end{equation}
Combining it with exact virial relation (\ref{inf}) for
$\,n=0\,$ and taking into account the equality\,
$\,1+\varkappa(\nu)\,=\,T\,(\partial
\nu/\partial \mathcal{P})_T\,$\, (see Section 3)
after clear reasonings one obtains
\begin{equation}
T\,\frac {\partial V_0(t,\Delta {\bf
R};\,\nu)}{\partial\mathcal{P}}\,+\, \overline{\Omega}(t,\Delta {\bf
R};\,\nu)\,\, V_0(t,\Delta {\bf R};\,\nu)\, \geq\,0\,\,\,,
\label{in1}
\end{equation}
where quantity\,
$\,\overline{\Omega}(t,\Delta {\bf R};\,\nu)\,$
is defined by
\begin{equation}
\overline{\Omega}(t,\Delta {\bf R};\,\nu)\,=\,\min_{0<\,\delta
<1}{\,\,\frac {1}{1-\delta} \int_{\Omega(\delta)} F_1^{(eq)}({\bf
r}|\,{\bf R};\nu)\,d{\bf r}}\, \,\approx\,\min_{0<\,\delta <1}{\frac
{\Omega(\delta,t,\Delta{\bf R};\nu)}{1-\delta}} \,\, \label{md}
\end{equation}
According to the aforesaid, if\, $\,V_1(t,{\bf R},{\bf r}|{\bf
R}_0;\,\nu)\,$ as a function of $\,{\bf r}$\, (at given other
arguments) has constant sign, then\, $\,\overline{\Omega}(t,\Delta
{\bf R};\,\nu)\,$ represents volume occupied by the pair correlation
or, briefly, ``pair correlation volume''. Otherwise
$\,\overline{\Omega}(t,\Delta {\bf R};\,\nu)\,$ is something smaller
than this volume.

Now discuss hypothetical asymptotic (\ref{vg})
from the viewpoint of inequality (\ref{in1}).
It says that the hypothesis can be true only if
\begin{equation}
\nu\,\overline{\Omega}(t,\Delta {\bf R};\,\nu)\,\geq\,-\,\nu T\,\frac
{\partial\ln D}{\partial \mathcal{P}}\left(\frac {\Delta {\bf
R}^2}{4Dt}-\frac 32\right)\,\rightarrow\,\frac {\Delta {\bf
R}^2}{4Dt}-\frac 32\,\,\,, \label{a}
\end{equation}
where last expression concerns dilute gas. In other words, if
quantity $\,\nu\,\overline{\Omega}(t,\Delta {\bf R};\,\nu)\,$,\, all
the more the pair correlation volume, measured in units of specific
volume per one atom, $\,1/\nu\,$, is not bounded above. In opposite,
if it is bounded above,
\begin{equation}
\nu\,\overline{\Omega}(t,\Delta {\bf
R};\,\nu)\,\leq\,c_1\,=\,\mathrm{const}\,\,\,, \label{bound}
\end{equation}
then Gaussian asymptotic (\ref{vg}) is forbidden. Instead, inequality
(\ref{in1}) allows for a generalized diffusion law as, for instance,
\begin{equation}
V_0(t,\Delta {\bf R};\,\nu)\,\rightarrow\,\frac {1}{(4\pi
Dt)^{3/2}}\,\Psi\left(\frac {\Delta {\bf R}^2}{4Dt}\right)\,
\Theta\left(\frac {|\Delta {\bf R}|}{v_0t}\right) \,\,\,, \label{as1}
\end{equation}
where function $\,\Psi(z)\,$ should satisfy inequality
\begin{equation}
z\,\frac {d\Psi(z)}{dz}\,+\,\alpha\,\Psi(z)\,\geq\,0\,\,\,,
\,\,\,\,\,\,\,\alpha\,\equiv\,\frac 32\,+\,c_1\left[-\,\nu T\,\frac
{\partial\ln D}{\partial \mathcal{P}}\right]^{-1}\,\,\label{in2}
\end{equation}
(supposing that $\,D\,$ falls when pressure grows). Consequently,
$\,\Psi(z\rightarrow\infty)\,\propto\,1/z^{\,\alpha}\,$\,. In the
case of gas, $\,\,\alpha =3/2+c_1\,$. Formula (\ref{as}) corresponds
to $\,c_1=\nu\overline{\Omega}=2\,$.

Hence, the theory inevitably leads to statistical correlations which
have unlimited extension either in space, as under variant (\ref{a}),
or in time, as under alternative variant (\ref{bound}). It remains to
ascertain what of the variants is closer to exact solution of BBGKY
equations. From physical point of view, the second one certainly is
preferable, at least if speak about gas. Indeed, as it is clear from
equations (\ref{vn}), a source of correlations between BP and atoms
is their collisions. A collision realizes at such disposition of BP
and atom when vector $\,{\bf r}-{\bf R}\,$ lies in the ``collision
cylinder'' which is oriented in parallel to the relative velocity
$\,{\bf p}/m-{\bf P}/M\,$ and has radius $\,\approx r_b\,$. At that
distance between colliding particles should not be much greater
than\, $\,\Lambda = \min {\,(\Lambda_a,\Lambda_b)}\,$, since
otherwise their collision almost surely will be prevented by an
encounter with the rest of atoms. Consequently, at any values of
momenta (all the more, at any $\,t\,$ and $\,\Delta {\bf R}\,$) it is
natural to estimate the pair correlation volume as\,
$\,\overline{\Omega}\,\approx\, 2\Lambda\, \pi r_b^2\,$\,. If BP is
mere marked atom \cite{p1}, then $\,\Lambda=\Lambda_b\,$\, and\,
$\,\overline{\Omega}\,\approx\, 2\Lambda_b\,\pi r_b^2\,=\,2/\nu\,$\,.
As the result, we arrive to (\ref{bound}) with $\,c_1\approx 2\,$.

\section{Conclusion}
Thus, to resume, the problem about thermodynamically equilibrium
random walk of a molecular ``Brownian'' particle (BP) in a fluid was
formulated as a problem of classical statistical mechanics of a
system of (infinitely) many particles. Corresponding BBGKY equations
for distribution functions (DF) and equation for their generating
functional were considered in terms of correlation functions (CF)
introduced so that they extract statistical correlations between
total path of BP during all its observation time and current state of
the medium (``historical correlations''). We showed (in case of Gibbs
canonical ensemble) that generating functionals of both equilibrium
DF and time-dependent CF are invariant in respect to definite
continuous group of transformations of their arguments including
density of the medium (mean concentration of particles).

The found invariance group produces a sequence of exact ``virial
relations'' which connect full sets of DF or CF taken at different
values of the density and therefore can serve as quality test of
approximate solutions of the BBGKY equations. In this respect, we
demonstrated that conventional ``Boltzmannian'' approximation to
kinetics of BP in dilute gas (or ideal gas) is incorrect. The matter
is that it rejects correlations of third and higher orders, while
actually correlations of any order are equally important, even (and
most of all) under the Boltzmann-Grad limit. The virial relations
automatically allow for all the correlations and therefore imply
significant restrictions on possible forms of probabilistic
distribution of the BP's path. In particular, they rather surely
(especially in case of dilute gas) forbid Gaussian asymptotic of this
distribution. What is possible instead of it is an automodel
diffusive asymptotic possessing power-law long tails (cut off at
ballistic flight length). This conclusion well agrees with previous
approximate solutions of BBGKY equations \cite{i1,i2,p1}. It means
that in real gas random BP's trajectories are so much unique that, in
contrast to the Lorentz gas, they can not be divided into
statistically independent constituent parts. In other words, they can
not be imitated by ``dice tosses''. In this sense actual molecular
random walk is non-ergodic
 \cite{bk3,i2,p1,bk12}.

In should be noted that results of this work can be easily extended
to thermodynamically non-equilibrium walk under influence of an
external force applied to BP from start of its observation (in this
respect see \cite{mpa3,jstat,i2}). From the other hand, it would be
interesting to generalize the mentioned invariance group to more
usual problems where not some select particles but hydrodynamic
fields are in the centre of attention.

I would like to acknowledge my colleagues
from DonPTI NANU Dr. I.\,Krasnyuk and Dr. Yu.\,Medevedev
for many useful discussions.

\end{document}